\documentclass[aps,prl,preprint]{revtex4}
\usepackage{graphicx,latexsym}

\begin{document}
 
\title{Inhomogeneous Fragmentation of the Rolling Tachyon}

\author{Gary Felder}
\affiliation{Smith College Physics Department, Northampton, MA, 01063,
  USA}
\email{gfelder@email.smith.edu}
\author{Lev Kofman}
\affiliation{CITA, University of Toronto, 60 St. George Street,
  Toronto, ON M5S 3H8, Canada}
\email{kofman@cita.utoronto.ca}
\date{\today}
\preprint{CITA-2004-8, hep-th/0403073}

\begin{abstract}
Dirac-Born-Infeld type effective actions reproduce many aspects of
string theory classical tachyon dynamics of unstable Dp-branes.  The
inhomogeneous tachyon field rolling from the top of its potential
forms topological defects of lower codimensions. In between them, as
we show, the tachyon energy density fragments into a p-dimensional
web-like high density network evolving with time. We present an
analytic asymptotic series solution of the non-linear equations for
the inhomogeneous tachyon and its stress energy. The generic solution
for a tachyon field with a runaway potential in arbitrary dimensions
is described by the free streaming of noninteracting massive particles
whose initial velocities are defined by the gradients of the initial
tachyon profile. Thus, relativistic particle mechanics is a dual
picture of the tachyon field effective action. Implications of this
picture for inflationary models with a decaying tachyon field are
discussed.
\end{abstract}

\maketitle

{\it Introduction.}
In this paper we investigate generic inhomogeneous solutions of
Dirac-Born-Infeld type theories
\begin{equation}\label{action}
S = -\int d^{p+1} x \, V(T) \sqrt{1 + \alpha' \partial_\mu T
\partial^\mu T} + {\cal O}\left( \partial_\mu \partial^\mu T \right) \
,
\end{equation}
where $T(x^{\mu})$ is a (dimensionless) scalar field and $V(T)$ is its
runaway potential (no minima). In string theory $\alpha'$ is a square
of the fundamental length scale; we put $\alpha'=1$. The action
(\ref{action}) was proposed in \cite{action} as an effective field
theory description of the open string theory tachyon which describes
unstable non-BPS D-branes. In application to the string theory tachyon
(\ref{action}) should be understood in the truncated approximation,
i.e. valid only in the regime where higher derivatives are not large.
The potential is often chosen to be $V(T)=\tau_p/{\cosh T}$ for the
bosonic case which we consider.  At large $T$ the potential has a
runaway character $V(T) \simeq e^{- T}$ with the ground state at
infinity.

There are several motivations for studying the properties of the
effective action (\ref{action}). It is difficult to find the open
string tachyon dynamics for generic tachyon inhomogeneities.  The
action (\ref{action}), meanwhile, permits us to study complicated
tachyon dynamics in terms of classical field theory. The relatively
simple formulation of tachyon dynamics in terms of the effective
action (\ref{action}) has therefore triggered significant interest in
the investigation of the field theory of the tachyon and the possible
role of tachyons in cosmology.  Indeed, the end point of string theory
brane inflation is annihilation of $ D-\bar D$ branes, which leads to
the formation and subsequent fragmentation of a tachyon condensate
\cite{infl}.  Thus the potential role of the tachyon in cosmology
cannot be understood without first understanding its fragmentation.

Apart from its application to the string theory tachyon, the search
for the structure of general solutions of the theory (\ref{action}) is
an interesting mathematical problem in and of itself. The equation of
motion arising from the action (\ref{action}) is an example of a
complicated, non-linear, partial differential equation which, as we
will show, admits a relatively simple, general, inhomogeneous
solution.  The evolution of the tachyon field $T(t, \vec x)$ can be
viewed as a mapping $T(t_0, \vec x_0) \to T(t, \vec x)$ that becomes
multivalued and generates singularities at caustics \cite{gks}.
Besides the DBI type theories (\ref{action}), there are other
cosmologically motivated phenomenological models of fields with high
derivatives which share the problematics of (\ref{action}).

If the potential $V(T)$ is symmetric around $T=0$ and the
inhomogeneous tachyon field begins rolling from the top of its
effective potential, then topological defects (kinks) can form due to
symmetry breaking \cite{kinks}.  In this paper we consider what
happens to the tachyon field in the region where it rolls down one
side of the potential.  We will see the formation of sharp features in
the tachyon energy density due to its fragmentation. These features,
which are related to the convergence of characteristics of the field
$T$, have to be distinguished from topological defects.  The full
picture must incorporate both effects, formation of kinks and tachyon
fragmentation in the space between them. Because it is hard to study
with CFT tachyon dynamics with a generic, spatially varying profile
\cite{sen}, previous calculations dealt with a plane wave tachyon
profile \cite{LNT}.  In this case the tachyon decays into equidistant
plane-parallel singular hypersurfaces of co-dimension one, which were
interpreted as kinks.  The effective action for a plane wave tachyon
predicts similar result, as we will see later. However, this
inhomogeneous profile is atypical in the sense that fragmentation
between kinks does not occur.  In the general case we expect both
structures, weblike fragmentation and topological defects.

\begin{figure*}
\includegraphics{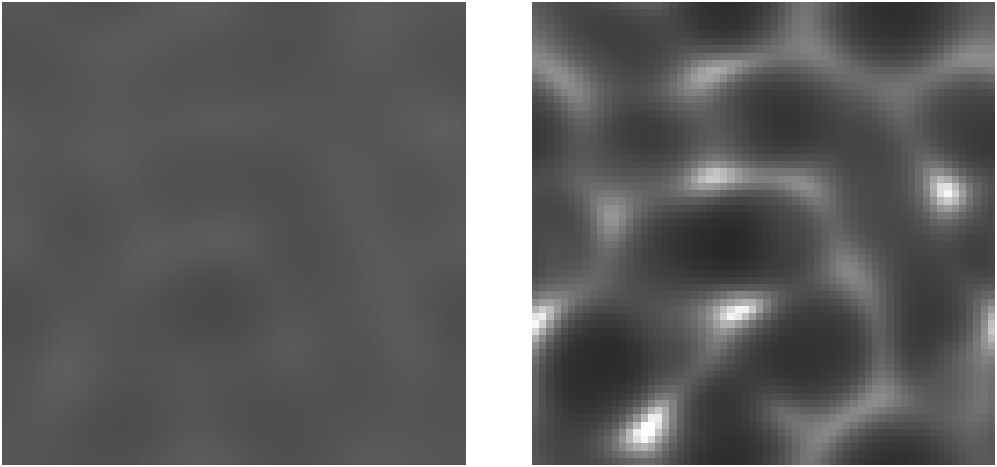}
\caption{\label{fig:web} The focusing of energy density into a
web-like structure due to caustic formation. Whiter regions correspond
to higher densities. The left panel shows a nearly homogeneous initial
Gaussian random field profile and the right panel shows the same field
a short time later.}
\end{figure*}

In this paper we concentrate on tachyon fragmentation between
kinks. We begin by showing an image that illustrates the fragmentation
of the tachyon field as it rolls down one side of its
potential. Figure \ref{fig:web} shows the result of a numerical
lattice simulation of the energy density of a two dimensional tachyon
field rolling down one side of its potential, as described by the
equation of motion (\ref{motion}) below. We used the LATTICEEASY code
\cite{latticeeasy} adapted for eq. (\ref{motion}). Starting from an
initial random Gaussian field $T$ the energy rapidly became fragmented
into an anisotropic structure of clumps joined by filaments into a
web-like network.

The tachyon energy density pattern in Fig. \ref{fig:web} is
reminiscent of the illumination pattern at the bottom of the swimming
pool or the web-like large scale structure of the universe. The
similarity is not coincidental: the underlying mathematics has common
features in all three cases. In what follows we present an analytic
solution to the tachyon equation of motion that describes in detail
the formation of this structure. We begin by describing a good
approximation to the dynamics of eq. (\ref{motion}) and go on to show
how to extend this approximation into an asymptotic series for the
field $T$.

{\it The Free Streaming Approximation.}
The equation of motion for the tachyon field follows from the action
(\ref{action})
\begin{equation}\label{motion}
\partial_\mu \partial^\mu T - {\partial_\mu \partial_\nu T \over 1 +
\partial_\alpha T \partial^\alpha T} \, \partial^\mu T \partial^\nu T
- {V_{,T} \over V} = 0 \ .
\end{equation}
For simplicity we will confine ourselves to a pure exponential
potential $V(T)=e^{-T}$, however our results are qualitatively valid
for any runaway potentials. The energy density of the tachyon field
$\rho=T_{00}$ is
\begin{equation}\label{rho}
\rho =\frac{e^{-T}}{\sqrt{1+\partial_{\mu}T \partial^{\mu}T}} \,
\dot{T}^2 + e^{-T} \, \sqrt{1+\partial_{\mu}T \partial^{\mu}T} \ ,
\end{equation}

We have observed solving (\ref{motion}) numerically \cite{gks} that if
we define an operator $P(T) = 1 + \partial_\mu T \partial^\mu T$ the
field $T$ rapidly approaches a regime in which $P(T) \approx 0$. We
write this by saying
\begin{equation}\label{zero}
T(x^{\mu}) \approx S(x^{\mu}) \ ,
\end{equation}
where $S$ satisfies the equation 
\begin{equation}\label{HJ}
\dot S^2-\left({\vec \nabla}_{x} S\right)^2=1 .
\end{equation}
The dot represents a time derivative and the spatial derivatives are
with respect to the $p$ spatial coordinates $\vec x$ on the
brane. This equation is the Hamilton-Jacobi equation for the evolution
of the wave front function of free steaming massive relativistic
particles.

Let us consider this particle description. At some initial time $t_0$
we can label the position of each particle with a vector
$\vec{q}$. Equivalently we can say that $\vec{q}$ parametrizes the
different particles. The initial four-velocity of the particle is
given by $\partial^{\mu} S_0$. If we further define the proper time
$\tau$ along each particle's trajectory, we can switch from
coordinates $(t, \vec{x})$ to $\left(\tau(t, \vec x), \vec q(t, \vec
x)\right)$ and obtain an exact parametric solution to (\ref{HJ})
\cite{gks}
\begin{eqnarray}\label{character}
\label{ssolutionx}\vec{x} &=& \vec{q} - \vec{\nabla}_{\vec q}S_0 \tau \\
\label{ssolutiont}t &=& \sqrt{1 + \vert \nabla_{\vec q}S_0 \vert^2} \, \tau \\
\label{ssolutions}S &=& S_0 + \tau.
\end{eqnarray}
The interpretation of the solution (\ref{character}) is very simple
and intuitive. It tells us that the field $S$ propagates along the
treajectories of the massive relativistic particles, growing linearly
with proper time. The slope of each characteristic depends only on the
initial gradients of $S_0$ on that characteristic.

In geometrical optics photons are massless, but the qualitative
picture of their wave front propagation is similar. This expains the
similarity between the two dimensional web-like pattern of Fig.
\ref{fig:web} and the illumination pattern at the bottom of a swimming
pool.  The focusing of particle trajectories corresponds to higher
density concentrations and further, to the formation of caustics at
the loci where trajectories cross.

{\it The Full Solution.} Let us now consider the energy density.
Looking at equation (\ref{rho}) we see that the exponential pieces are
growing exponentially small, as are the arguments of the square
roots. The second term in (\ref{rho}) will thus rapidly become
irrelevant and we need consider only how the exponentially small
numerator and denominator of the first term will be related. (The
leading term in $\dot{T}^2$ is one).

To calculate $\rho$ we will need to go beyond the free streaming
approximation (\ref{zero}). In view of the exponential in the
numerator of $\rho$ we conjecture that $T$ can be expanded as
\begin{equation}\label{first}
T(x^{\mu}) \approx S + f_1[S] \, e^{-2S} \ ,
\end{equation}
where $f_1$ is a sub-exponential functional of $S$. (We could include
a lower order term $f_0 e^{-S}$. As we explain below, we can solve
exactly for $f_0$ in that case and we find that its effects can be
absorbed into $S$ and $f_1$.) Now we are going to check the validity
of this expansion.

Plugging the expansion (\ref{first}) into equation (\ref{motion}) and
keeping only terms proportional to $e^{-2 S}$ gives the following
equation for the sub-exponential function $f_1$
\begin{equation}\label{f1motion}
\left(\partial^\mu S \partial^\nu S\right) \, \partial_\mu
\partial_\nu f_1 + 2 \left(1 - \Box S\right) \partial^\mu \, S
\partial_\mu f_1 - 4 \Box S \, f_1 = 0 \ ,
\end{equation}
where $\Box S=\partial_\nu \partial^\nu S=-\ddot S + {\nabla_{\vec
x}}^2 S$.

This equation can be dramatically simplified by changing from ($t$,
$\vec{x}$) coordinates to ($\tau$, $\vec{q}$) coordinates, as defined
by the characteristics of $S$ in (\ref{ssolutionx}-\ref{ssolutiont}).
In these coordinates, $f_1$ has no spatial derivatives and equation
(\ref{f1motion}) reduces to
\begin{equation}\label{f1}
f_{1,\tau\tau} - 2 \left(1 - \Box S\right) f_{1,\tau} - 4 \Box S \,
f_1 = 0,
\end{equation}
Note that $\Box S$ can be calculated either with respect to ($t$,
$\vec{x}$) or ($\tau$, $\vec{q}$) coordinates. We can further simplify
this equation by introducing a new variable $y$
\begin{equation}\label{y}
y \equiv 2 f_1 - f_{1,\tau} \ .
\end{equation}
Equation (\ref{f1}) can be rewritten in terms of  $y$ 
\begin{equation}
y_{,\tau} + 2 \Box S \, y = 0,
\end{equation}
which can be immediately solved to give $y(\tau, \vec q) = y_0(\vec q)
\exp\left(-2 \int^{\tau} d\tau' \, \Box S \right) $. From this and
(\ref{y})
\begin{equation}\label{fone}
f_1(\tau, \vec q)=f_{1i}(\vec q) \, e^{2\tau} \,
\int^{\tau}d\tau'e^{-2\tau' -2 \int^{\tau'} d\tau'' \, \Box S} \ .
\end{equation}

To proceed further we need to calculate $\Box S$.  In principle it can
be done from the solution (\ref{character}) in parametric form by
inverting the $(t, \vec{x}) \to (\tau, \vec{q})$ coordinate
transformation matrix \footnote{We have carried out this explicit
calculation in one and two dimensions as a check on the method
described below.}.  Instead we will use the following trick. Plugging
the expansion (\ref{first}) into the energy density (\ref{rho}) we
find $\rho \approx 1/\sqrt{2\left(2f_1+\partial_\mu S \partial^\mu f_1
\right)}$.  Observe that the denominator here is {\it exactly
identical} to $\sqrt{2y}$, thus
\begin{equation}\label{energy}
\rho \approx {1 \over \sqrt{2 y}} = \rho_0(\vec q) \, e^{ \int^{\tau}
d\tau' \Box S } \ .
\end{equation}
This is precisely the expression for the energy density of free
streaming relativistic particles which obey the relativistic
continuity equation $\partial_\mu S \left( \rho \, \partial^\mu S
\right)=0$. In fact, (\ref{energy}) is the solution of this continuity
equation in the coordinates $(\tau, \vec{q})$.  However, there is
another form of the energy density, which is equivalent to
(\ref{energy})
\begin{equation}\label{energy1}
\rho ={{\rho_0(\vec q)} \over {\vert \frac{\partial \vec x}{ \partial
\vec q } \vert }} \ ,
\end{equation}
where the denominator is the Jacobian of the $\vec x \to \vec q$
transformation.  Indeed, from conservation of the energy density in a
differential volume we have $d^p \vec q \rho_0(\vec q)=d^p \vec x
\rho$, which leads to the formula (\ref{energy1}).  It is now
straightforward to calculate the Jacobian from the formulas
(\ref{ssolutionx}-\ref{ssolutiont}).  In $p$ dimensions it is a
polynomial in $\tau$ of order $p$.  For example, in the one
dimensional case $ \vert \frac{\partial \vec x}{ \partial \vec q }
\vert =(\tau_{c} - \tau)$, where $\tau_{c}(\vec q)$ is a function of
the gradients of $S_0$.  In the two dimensional case $ \vert
\frac{\partial \vec x}{ \partial \vec q }\vert = (\tau_{c1} - \tau)
(\tau_{c2} - \tau) $, where $\tau_{cn}$ are function of $\nabla_{\vec
q} S_0$.

Comparing expressions (\ref{energy}) and (\ref{energy1}), we can find
$\Box S$ and calculate $f_1$ using (\ref{fone}).  We find that a
constant originating from the integration in (\ref{fone}) can be
absorbed in $S$ while the rest of $f_1$ will be a $2 p$ order
polynomial in $\tau$. This demonstrates the validity of the
approximation (\ref{first}) by showing that the $f_1$ term provides
only exponentially suppressed corrections to the leading term.

We could go further and include other powers of $e^{-S}$ in our
expansion (\ref{first}).
\begin{equation}\label{full}
T(x^{\mu}) \approx S+ \sum_{n=0}^{\infty}f_n[S] \, e^{-(n+1)S} \ .
\end{equation}
We have explicitly checked all such terms up through $e^{-4 S}$,
including a possible term proportional to $e^{-S}$, and found that
they simply provide corrections to the integration constants of $S$
and $f_1$ plus terms that are exponentially suppressed relative to the
ones we have discussed. We can further show that all such terms $f_n$
have the same characteristics as $S$, and we therefore conclude that
the general inhomogeneous solution $T$ propagates along the
characteristics (\ref{ssolutionx}-\ref{ssolutiont}). Up to
exponentially small corrections (which could in principle be
calculated order by order), the complete solution for $T$ can be
represented by the two functions $S_0(\vec q)$ and $f_{1i}(\vec q)$.

With these results, we can evaluate the energy density (\ref{energy1})
to leading order using only $f_1$.  For an arbitrary brane dimension
$p$ we have
\begin{equation}
\rho \approx \rho_0 \,  \prod_{n=1}^{p} (1-\lambda_n t)^{-1} \ ,
\end{equation}
where $\lambda_n(\vec{q})$ are the eigenvalues of $\partial_\mu
\left(\dot{S}_0^{-1} \partial_\nu S_0\right)$. From here we see that
the energy density first reaches large values in regions where
$\lambda_n(\vec{q})$ is maximal. For some critical trajectories $\vec
q_c$ at a critical time $t_c$ the energy density becomes singular,
which corresponds to caustic formation.  This is exactly what one
would expect in the picture of free streaming, massive, relativistic
particles, where the energy density blows up at the orbit crossings.

{\it Conclusions.}
Our most important conclusion is that the general inhomogeneous
solution of the field theory (\ref{action}) very rapidly approaches
the asymptotic form (\ref{full}), which is equivalent to the
relativistic mechanics of freely propagating massive particles with
velocities $\vec v_a=-\nabla_{\vec q} S_0(\vec q_a)$. In other words,
there is a duality between the two Lagrangians
\begin{equation}\label{dual}
 V(T) \sqrt{1 + \partial_\mu T \partial^\mu T} \Longleftrightarrow
 \sum_a \sqrt{1-\vec v_a^2} \ .
\end{equation}
The whole process of unstable $Dp$ brane decay in the the dual picture
is described as ``crumbling to dust'' of massive particles.  A
complete discussion of the interpretation of massive particles and
anisotropic high density structures (clumps, filaments, sheets) which
they form would be beyond the scope of this paper.  In the $1+1$ case
the high density regions of the orbit crossing were conjectured to be
$D0$ branes \cite{BCKR}.

From the velocities of the characteristics (\ref{ssolutionx}) we can
see that around maxima of the initial field profile the
characteristics will tend to diverge and the profile will flatten. In
regions around minima, however, characteristics will tend to converge,
the profile will become sharper, and after some critical time $t_c$
the field solution will become multivalued. In the dual picture of
relativistic particles this corresponds to caustic formation. At the
caustics the energy density blows up. Caustic formation also entails
divergences in the second derivatives of $T$, which signal the
breakdown of the truncated approximation (\ref{action}). In short the
Lagrangians (\ref{dual}) are unable to describe the field $T$ when its
solution becomes multi-valued. In the picture of freely moving massive
particles we can inlcude interactions to cause them to stick together
as their trajectories intersect with impact parameter $ \sim
\sqrt{\alpha'}$.

Let us make a remark about a one-dimensional plane wave tachyon
profile $T = \cos(x)$ rolling from the top of its (symmetric)
effective potential. Since the parts of the field that roll to the
right have no minima they do not form caustics, and by symmetry the
parts rolling to the left do not either. In this particular case the
tachyon fragments into kinks only.

We can also consider a more general profile, however. Tachyonic
instability occurs for all inhomogeneous modes $k$ for which the
effective mass $m^2=k^2 -1/\alpha'^2$ is negative.  Therefore the
generic tachyon initial profile is a superposition of a number of
modes, which produces a random Gaussian field $T_0(\vec{q})$. These
initial conditions typically arise from quantum fluctuations during
symmetry breaking (see e.g. \cite{SB}). In this case we once again
expect the formation of topological defects due to symmetry
breaking. Outside these defects, however, the tachyon field will
fragment into a web like structure as shown in Fig. \ref{fig:web}. If
the defects are walls these webs will form within each domain; if they
are strings the web of caustics will be mixed in with the strings.

If the spectrum of $T_0(\vec q)$ inhomogeneities has scaling
properties (as quantum fluctuations do), then the web-like network
will evolve in a scaling manner. The smallest scale of the web is
defined by the largest tachyonic mode $k$. The dual picture of freely
moving massive particles which stick together as their orbits
intersect gives a simple explanation of such fragmentation.

Finally, we note the relevance of our result for cosmological
applications of the tachyon.  In the context of brane inflation, which
ends with a pair of $D3-\bar D3$ branes annihilating, the tachyon is a
complex field and strings will be created and the rst of energy is
transform into radiation \cite{Jones}.  If the result of real tachyon
field fragmentation which we derive above is extended to the complex
tachyon, then annihilation of $D3-\bar D3$ branes results in the net
of strings plus massive particles with the matter dominant equation of
state. The absence of radiation domination after brane inflation may
pose a problem for the model \cite{KL}

\begin{acknowledgments}
We are grateful to  D. Kutasov, A. Linde, R. Myers,  S. Shandarin, and
J. Martin for useful discussions. The work by L.K. was supported by
NSERC and CIAR. G.F. wishes to thank CITA for its hospitality during
this research..
\end{acknowledgments}


\begin{thebibliography}{99}

\bibitem{action}
A.~Sen,``Rolling tachyon,'' JHEP 0204, 048 (2002),
arXiv:hep-th/0203211;
 A.~Sen,  ``Tachyon matter,''  JHEP 0207, 065 (2002)
arXiv:hep-th/0203265;
  A.~Sen, ``Field theory of tachyon matter,''
 Mod. Phys. Lett. A17, 1797 (2002), arXiv:hep-th/0204143; 
J. Kluson, ``Proposal for non-BPS D-brane action'',
Phys. Rev. D62, 126003 (2000), arXiv:hep-th/0004106;
M. Garousi, ``Tachyon couplings on non-BSP D-branes and Dirac-Born-Infeld
action'' Nucl. Phys. B584, 284 (2000), arXiv:hep-th/0003122;
J.~Minahan, ``Rolling the tachyon in super BSFT'', JHEP 0207, 030 (2002)
 arXiv:hep-th/0205098; E. Bergshoeff, M. de Roo, T. de Wit, E. Eyras
and S. Panda, ``T-duality and actions for non-BPS D-branes,  JHEP 0005, 009 (2000),
arXiv:hep-th/0003221;
D.~Kutasov and V.~Niarchos,
``Tachyon effective actions in open string theory,''
arXiv:hep-th/0304045.

\bibitem{infl} C.~P.~Burgess, M.~Majumdar, D.~Nolte, F.~Quevedo, G.~Rajesh and
R.~J.~Zhang, ``The inflationary brane-antibrane universe,'' JHEP
{\bf 0107}, 047 (2001) [arXiv:hep-th/0105204];
N.~Jones, H.~Stoica and S.~H.~Tye,
``Brane interaction as the origin of inflation,''
arXiv:hep-th/0203163;
S.~Kachru, R.~Kallosh, A.~Linde, J.~Maldacena, L.~McAllister and S.~P.~Trivedi,
JCAP {\bf 0310}, 013 (2003)
[arXiv:hep-th/0308055].

\bibitem{gks} G.~N.~Felder, L.~Kofman and A.~Starobinsky,
``Caustics in tachyon matter and other Born-Infeld scalars,'' JHEP
{\bf 0209}, 026 (2002) [arXiv:hep-th/0208019].


\bibitem{kinks} G.~Shiu, S.~H.~H.~Tye and I.~Wasserman, ``Rolling
tachyon in brane world cosmology from superstring field theory,''
Phys.\ Rev.\ D {\bf 67}, 083517 (2003) [arXiv:hep-th/0207119];
J. Cline, H. Firouzjahi and P. Martineau, hep-th/0207156; A. Sen,
``DBI action on tachyon kink and vortex'' [arXiv:hep-th/0303057].

\bibitem{sen} 
A.~Sen,``Time evolution in open string theory,''
JHEP {\bf 0210}, 003 (2002)
[arXiv:hep-th/0207105].


\bibitem{LNT}
J.~A.~Harvey, D.~Kutasov and E.~J.~Martinec,
``On the relevance of tachyons,''
arXiv:hep-th/0003101;
F.~Larsen, A.~Naqvi and S.~Terashima,
``Rolling tachyons and decaying branes,''
JHEP {\bf 0302}, 039 (2003)
[arXiv:hep-th/0212248].

\bibitem{latticeeasy} G.~N.~Felder and I.~Tkachev, ``LATTICEEASY: A
program for lattice simulations of scalar fields in an expanding
universe,'' [arXiv:hep-ph/0011159].

\bibitem{BCKR}
M.~Berkooz, B.~Craps, D.~Kutasov and G.~Rajesh,
``Comments on cosmological singularities in string theory,''
JHEP {\bf 0303}, 031 (2003)
[arXiv:hep-th/0212215].

\bibitem{SB}
G.~N.~Felder, L.~Kofman and A.~D.~Linde,
``Tachyonic instability and dynamics of spontaneous symmetry breaking,''
Phys.\ Rev.\ D {\bf 64}, 123517 (2001)
[arXiv:hep-th/0106179].

\bibitem{Jones}
N.~T.~Jones, H.~Stoica and S.~H.~H.~Tye,
 ``The production, spectrum and evolution of cosmic strings in brane
inflation,''
Phys.\ Lett.\ B {\bf 563}, 6 (2003)
[arXiv:hep-th/0303269].


\bibitem{KL}
L.~Kofman and A.~Linde,
``Problems with tachyon inflation,''
JHEP {\bf 0207}, 004 (2002)
[arXiv:hep-th/0205121].


\end{thebibliography}
\end{document}